\documentclass[twocolumn,aps,pra,showpacs,amsfonts]{revtex4}

\def\endproof{\vrule height6pt width6pt depth0pt}

\begin{document}
\title{Finite-precision measurement does not nullify
the Kochen-Specker theorem}
\author{Ad\'{a}n Cabello}
\email{adan@us.es}
\affiliation{Departamento de F\'{\i}sica
Aplicada II, Universidad de Sevilla, 41012 Sevilla, Spain}
\date{\today}


\begin{abstract}
It is proven that any hidden variable theory of the type proposed
by Meyer [Phys. Rev. Lett. {\bf 83}, 3751 (1999)], Kent [{\em
ibid.} {\bf 83}, 3755 (1999)], and Clifton and Kent [Proc. R. Soc.
London, Ser. A {\bf 456}, 2101 (2000)] leads to experimentally
testable predictions that are in contradiction with those of
quantum mechanics. Therefore, it is argued that the existence of
dense Kochen-Specker-colorable sets must not be interpreted as a
nullification of the physical impact of the Kochen-Specker theorem
once the finite precision of real measurements is taken into
account.
\end{abstract}



\maketitle


\section{The Kochen-Specker theorem}
The Kochen-Specker (KS) theorem \cite{Specker60,Bell66,KS67}
shows one of the most fundamental features of quantum mechanics
(QM): measurements do not reveal preexisting values.
More precisely, it asserts that any hidden variable theory that
satisfies QM must be {\em contextual} (i.e., the predefined
results must change depending on which other compatible measurements
are performed).

Its original mathematical proof \cite{KS67}
is based on the observation that, for a physical
system described in QM by a
Hilbert space of dimension $d \ge 3$,
it is possible to find a set of $n$ projection operators $P_i$
which represent yes-no questions about the physical system
so that none of the $2^n$ possible sets of ``yes'' or ``no'' answers
is compatible with the sum rule of QM for orthogonal resolutions of
the identity (i.e., if the sum of a subset of mutually orthogonal
projection operators is the identity, one and only one of the
corresponding answers ought to be ``yes'') \cite{Peres93}.
Yes-no questions can also be represented by the vectors $\hat{v}_i$
onto which $P_i$ projects.
$\hat{v}_i$ can be assumed to belong to
$S^{d-1}$, the unit sphere in ${\mathbb R}^d$. If there are predefined
noncontextual yes-no answers, then there will exist a function
$f:S^{d-1} \longrightarrow \{0,1\}$
such that
\begin{equation}
\sum_{i=1}^{d} f(\hat{v}_i)=1\;\;\mbox{whenever}\;\;
\sum_{i=1}^{d} P_i=1\!\!\:\!{\rm{I}},
\label{sumrule}
\end{equation}
where $\{ P_i \}_{i=1}^d$ is a set of orthogonal projectors and
$1\!\!\:\!{\rm{I}}$ denotes the identity, $f(\hat{v}_i)=1$ means
that the predefined answer to the yes-no question represented by
$P_i$ is ``yes,'' and $f(\hat{v}_j)=0$ means that the answer to
$P_j$ is ``no.'' If such a function exists for a given set of
vectors, it is said that the set is ``KS-colorable;'' if it does
not exist, then it is said that the set is ``KS-uncolorable'' and
serves as a proof of the KS theorem. The original proof
\cite{KS67} consists of a KS-uncolorable set of 117 vectors in
$S^2$. The smallest proofs currently known have $31$ vectors in
$S^2$ \cite{CK90} and $18$ vectors in $S^3$ \cite{CEG96}.

A simple physical interpretation of the projection operator $P_i$
onto $\hat{v}_i \in S^2$ can be given in terms of the spin
components along $\hat{v}_i$ of a spin-1 particle. Using a
suitable representation for $J_x$, $J_y$, and $J_z$ \cite{com1},
the relation
\begin{equation}
P_i=1\!\!\:\!{\rm{I}}-{J_i^2 \over \hbar^2}
\label{relation}
\end{equation}
defines a one-to-one correspondence between the projector $P_i$
onto $\hat{v}_i$ and the square of the spin component along
$\hat{v}_i$, denoted by $J_i^2$. Therefore, a measurement of $P_i$
represents the yes-no question ``does the square of the spin
component along $\hat{v}_i$ equal zero?'' The eigenvalue 1
corresponds to the answer ``yes,'' and the degenerate eigenvalues
0 to the answer ``no.'' The operators $J_x^2$, $J_y^2$, $J_z^2$
(or any other three squares of spin components along three
orthogonal directions) commute, so that the corresponding
observables can be measured simultaneously. In addition, since
\begin{equation}
J_x^2+J_y^2+J_z^2=2 \hbar^2 1\!\!\:\!{\rm{I}},
\label{squareofspincomponents}
\end{equation}
then QM predicts that the results of measuring observables
$J_x^2$, $J_y^2$, $J_z^2$ must be one 0 and two $\hbar^2$.
Analogously, the projectors $P_x$, $P_y$, $P_z$ commute, so that
the corresponding yes-no questions can be measured simultaneously.
Using Eq.~(\ref{relation}), Eq.~(\ref{squareofspincomponents})
becomes
\begin{equation}
P_x+P_y+P_z=1\!\!\:\!{\rm{I}}.
\label{projectors}
\end{equation}
Therefore, according to QM, the answers to $P_x$, $P_y$, $P_z$
must be one ``yes'' (represented by 1) and two ``no'' (represented by 0).

The fact that a function $f:S^2 \longrightarrow \{0,1\}$ satisfying the condition
(\ref{sumrule}) does not exist means that all $P_i$
($J_i^2$) cannot have predefined answers (values) compatible with
relation (\ref{projectors}) [(\ref{squareofspincomponents})].


\section{Proofs of ``nullification''}
Godsil and Zaks \cite{GZ88} have shown that the three-dimensional
rational unit sphere $S^2 \cap {\mathbb{Q}}^3$ can be colored
using only three colors such that orthogonal vectors are
differently colored. A corollary of this result has been recently
used by Meyer \cite{Meyer99} to show that $S^2 \cap
{\mathbb{Q}}^3$ is KS-colorable. Therefore, one can assign
predefined answers if one is restricted to those $P_i$ which
project onto $\hat{v}_i \in S^2 \cap {\mathbb{Q}}^3$. The set
$S^2 \cap {\mathbb{Q}}^3$ is dense in $S^2$ and therefore vectors
in $S^2 \cap {\mathbb{Q}}^3$ cannot be distinguished from those in
$S^2$ by finite-precision measurements. This leads Meyer to
conclude that ``finite-precision measurement nullifies the
Kochen-Specker theorem'' \cite{Meyer99}.

Kent \cite{Kent99} has
shown that dense KS-colorable sets exist in any arbitrary
finite-dimensional real or complex Hilbert space. This leads him to
conclude that ``noncontextual hidden variable (NCHV) theories
cannot be excluded by theoretical arguments of the KS type once
the imprecision in real world measurements is taken into account''
\cite{Kent99}. More recently, Clifton and Kent \cite{CK00} have
constructed a NCHV model for any finite-dimensional Hilbert space
that they claim is consistent with all the {\em statistical}
predictions of QM.
This allows them to conclude that
``all the predictions of nonrelativistic QM
that are verifiable to within any finite precision {\em can} be
simulated classically by NCHV theories'' \cite{CK00}.


In response, Ax and Kochen \cite{AK99} have argued that the study
of the effect of finite-precision measurements on the KS theorem
requires a different formalization which is still missing. In
\cite{Cabello99} there is a criticism of the physical
interpretation of the existence of KS-colorable sets. Havlicek
{\em et al.} \cite{HKSS99} have argued that any possible
KS coloring of the rational unit sphere is not physically
satisfactory. Mermin \cite{Mermin99} has argued that the
continuity of probabilities under slight changes in the
experimental configuration weighs against the conclusions in
\cite{Meyer99,Kent99}. Appleby has expanded on Mermin's discussion
\cite{Appleby00a} and argued that in the models of Meyer, Kent,
and Clifton the very existence of an observable is contextual
\cite{Appleby00b} and measurements do not reveal preexisting
classical information \cite{Appleby01}.

In this paper I shall prove that any possible KS coloring of the
rational unit sphere of the type proposed in \cite{Meyer99} leads
to predictions which differ from those of QM and, therefore, that
any NCHV theory which assigns definite colors to the rational unit
sphere can be discarded on experimental grounds even if
finite-precision measurements are used. In addition, I shall prove
that any possible KS coloring of a dense set of the kind proposed
by Clifton and Kent \cite{CK00} also leads to predictions that
differ from those of QM, and therefore explicit NCHV models like
those in \cite{CK00} can also be discarded on experimental
grounds.


Both proofs are inspired by a lesser known type of
proof of the KS theorem which does
not require an entire KS-uncolorable set but only
one of its subsets \cite{CG96}. In particular, it is based on
a proof by Stairs \cite{Stairs83}
which only requires eight of the vectors of
the 117-vector KS-uncolorable set in \cite{KS67}.
This eight-vector set appears for the first time in \cite{KS65}.
Stairs' proof was reformulated by Clifton \cite{Clifton93}
(see also \cite{Helle94,Vermaas94,CG95}).

The strategy of both proofs is as follows.
First we show that any NCHV theory cannot assign
the value 1 to certain vectors in the set
which is assumed to have predefined values.
We then show that the impossibility of such an assignment
leads to an inequality valid for any
NCHV theory but which is violated by QM.


\section{KS colorings of the rational sphere are incompatible with QM}


Consider the following vectors of the rational unit sphere:
\begin{eqnarray}
\hat{A} & = & \left({0,c_A,-s_A}\right),
\label{uno} \\
\hat{B} & = & \left({c_B,s_B,0}\right), \\
\hat{C} & = & (c_C c_D+ N c_A s_C s_D,
N s_A s_C c_D, \nonumber \\
 & & -c_C s_D+ N c_A s_C c_D ),
\end{eqnarray}
where $s_i=\left({1-c_i^2}\right)^{1/2}$,
$i$ being $A$, $B$, $C$, or $D$,
$N=\left[c_A^2+(s_A c_D)^2\right]^{-1/2}$,
and $\{c_i, s_i, N\} \in {\mathbb{Q}}$.

{\em Lemma 1.}
There is no KS coloring (\ref{sumrule})
of the rational unit sphere
$S^2 \cap {\mathbb{Q}}^3$ in which
$f(\hat{A})=f(\hat{B})=f(\hat{C})=1$.

{\em Proof via a reductio ad absurdum}.
Consider the following additional vectors
$\hat{v}_i \in S^2 \cap {\mathbb{Q}}^3$:
\begin{eqnarray}
\hat{v}_1 & = & \left({1,0,0}\right), \\
\hat{v}_2 & = & N \left({c_A s_D,s_A c_D,c_A c_D}\right), \\
\hat{v}_3 & = & \left({0,0,1}\right), \\
\hat{v}_4 & = & N \left({s_A c_D s_D,-c_A,s_A c_D^2}\right), \\
\hat{v}_5 & = & \left({0,1,0}\right), \\
\hat{v}_6 & = & \left({c_D,0,-s_D}\right).
\label{nueve}
\end{eqnarray}
If $f(\hat{A})=1 \Rightarrow f(\hat{v}_1)= f(\hat{v}_2) =0$,
if $f(\hat{B})=1 \Rightarrow f(\hat{v}_3)=0$, and
if $f(\hat{C})=1 \Rightarrow f(\hat{v}_4)=0$.
In addition,
if $f(\hat{v}_1)=f(\hat{v}_3)=0 \Rightarrow f(\hat{v}_5)=1$ and
if $f(\hat{v}_2)=f(\hat{v}_4)=0 \Rightarrow f(\hat{v}_6)=1$.
However, $f(\hat{v}_5)=1$ is incompatible with
$f(\hat{v}_6)=1$ since $\hat{v}_5$ and $\hat{v}_6$ are orthogonal.
\hfill\endproof


Let us consider an ensemble of spin-1 particles
and let us assume that any particle has a definite
color (1 or 0) for every vector of the rational unit sphere
and in particular for $\hat{A}$, $\hat{B}$,
and $\hat{C}$.
Let $P(B)$ be the probability of finding a particle
with $f(\hat{B})=1$ in such an ensemble,
$P(A \wedge B \wedge C)$ be the probability of
finding $f(\hat{A})=f(\hat{B})=f(\hat{C})=1$,
$P(A \wedge B \wedge \neg C)$ be the probability of
finding $f(\hat{A})=f(\hat{B})=1$ and
$f(\hat{C})=0$, and
$P(A|B)$ be the probability of finding $f(\hat{A})=1$
if $f(\hat{B})=1$.
Note that such probabilities make sense in
a NCHV theory
but not in QM.
From the point of view of a NCHV theory,
$P(A|B)$ can be obtained by means of
two alternative but equivalent methods \cite{Clifton93,CG95}:
either preparing the particles in a quantum eigenstate of
the square of the spin component along
$\hat{B}$ with eigenvalue 0 \cite{com1}, measuring the
square of the spin along $\hat{A}$,
and counting the number of events in which the eigenvalue 0 has
been obtained, or
preparing the particles in pairs in the singlet state,
measuring the square of the spin along $\hat{B}$
in one of the particles and the square of the spin
along $\hat{A}$ in the other, and counting the number
of events in which both results are 0.


{\em Lemma 2.}
There is no KS coloring (\ref{sumrule})
of the rational unit sphere $S^2 \cap {\mathbb{Q}}^3$ compatible
with all the statistical predictions of QM.

{\em Proof.}
The following inequality must be satisfied
in any NCHV theory:
\begin{equation}
P(B) \ge P(A \wedge B \wedge \neg C)+P(\neg A \wedge B \wedge C).
\label{ine1}
\end{equation}
Lemma 1 shows that
\begin{equation}
P(A\wedge B \wedge C)=0.
\end{equation}
Therefore, the inequality (\ref{ine1}) can be written as
\begin{equation}
P(B)\ge P(A \wedge B)+P(B \wedge C),
\label{ine2}
\end{equation}
which is equivalent to
\begin{equation}
P(B) \ge P(B) P(A|B)+P(B) P(C|B).
\label{ine2d}
\end{equation}
Inequality (\ref{ine2d}) can be simplified to
\begin{equation}
1 \ge P(A|B)+P(C|B).
\label{ine2e}
\end{equation}
Let us define
\begin{equation}
F_{\rm{NCHV}}=P(A|B)+P(C|B).
\label{F}
\end{equation}
Then, according to inequality
(\ref{ine2e}), any NCHV theory will predict
\begin{equation}
F_{\rm{NCHV}} \le 1.
\label{NCHV}
\end{equation}

On the other hand, in QM the equivalent of Eq.~(\ref{F}) is
\begin{equation}
F_{\rm{QM}}= \left|{\langle {\hat{A}|\hat{B}} \rangle}\right|^2+
\left|{\langle {\hat{C}|\hat{B}} \rangle}\right|^2.
\end{equation}
In particular, if we choose
\begin{eqnarray}
c_A & = & \frac{104}{185}, \\
c_B & = & \frac{10209400000}{12605796209}, \\
c_C & = & \frac{490231}{789769}, \\
c_D & = & \frac{105}{137},
\end{eqnarray}
then we obtain
\begin{equation}
F_{\rm{QM}} = 1.108,
\label{QM}
\end{equation}
which contradicts the prediction of NCHV theories
given by inequality (\ref{NCHV}) \cite{com2}.
\hfill\endproof


If all the particles of the ensemble have predefined ``colors''
along $\hat{A}$, $\hat{B}$, and $\hat{C}$, then $F$ has an
``exact'' value for that ensemble. To check that value, we must
perform tests along $\hat{A}$ and $\hat{B}$, and along $\hat{C}$
and $\hat{B}$. When we perform tests along $\hat{A}$ and
$\hat{B}$, their results reveal either the real colors of
$\hat{A}$ and $\hat{B}$ or the colors of $\hat{A}'$ and $\hat{B}'$
that are, respectively, infinitesimally close to them. In any NCHV
theory in which measurements are assumed to reveal predefined
colors and because of the own definition of ``precision,'' in a
higher precision test along $\hat{A}$ and $\hat{B}$, the number of
results revealing the true colors of $\hat{A}$ and $\hat{B}$ is
higher than in a lower precision test. Therefore, successive tests
with increasing precision will give us the true colors with a
higher probability. Thus, they will give us decreasing bounds
around the exact value of $F$. Therefore, even experiments with
finite-precision can discriminate between the prediction of NCHV
theories (\ref{NCHV}) and the prediction of QM (\ref{QM}).


\section{Clifton and Kent's NCHV model is incompatible with QM}


Clifton and Kent's NCHV model \cite{CK00} is based on the existence of dense
KS-colorable sets ${\cal D}$ with the remarkable property that
every projector in ${\cal D}$ belongs to only one resolution of the identity.
Moreover, the function $f$ defined over ${\cal D}$
and satisfying condition (\ref{sumrule})
must be ``sufficiently rich to recover the statistics of any
quantum state'' \cite{CK00}.
They claim that the existence of ${\cal D}$ ``defeats the practical
possibility of falsifying NCHV on either nonstatistical or
statistical grounds'' \cite{CK00}.

Let us outline a proof that shows that such a claim is not correct.
Consider a particular ${\cal D}$ dense in $S^2$
and two vectors $\hat{A}',\hat{B}' \in {\cal D}$.
Suppose that $\hat{A}'$ is infinitesimally close to
$\frac{1}{\sqrt{3}}(1,1,1)$
and $\hat{B}'$ is infinitesimally close to
$\frac{1}{\sqrt{3}}(1,1,-1)$.


{\em Lemma 3.}
Given an ensemble of systems
such that each system is described by ${\cal D}$,
the probability of finding an individual system in which
a KS coloring (\ref{sumrule})
satisfies
$f(\hat{A}')=f(\hat{B}')=1$ is infinitesimally close to zero
if such coloring must simulate the predictions of QM.

{\em Proof.}
Consider six additional vectors
$\hat{v}'_i \in {\cal D}$ such that
$\hat{v}'_1$ and $\hat{v}'_2$ are
both infinitesimally close to being orthogonal to $\hat{A}'$;
$\hat{v}'_3$ and $\hat{v}'_4$ are
both infinitesimally close to being orthogonal to $\hat{B}'$;
$\hat{v}'_1$, $\hat{v}'_3$, and $\hat{v}'_5$ are mutually
orthogonal;
$\hat{v}'_2$, $\hat{v}'_4$, and $\hat{v}'_6$ are mutually
orthogonal;
$\hat{v}'_5$ and $\hat{v}'_6$ are infinitesimally
close to being orthogonal \cite{com5}.
The fact that eight vectors with the above properties exist in ${\cal D} \in S^2$
is not excluded in \cite{CK00}.
Since $f$ must simulate the predictions of QM,
if $f(\hat{A}')=1$, then the probability of $f(\hat{v}'_1)=1$
must be $\left|{\langle {\hat{v}'_1|\hat{A}'} \rangle}\right|^2$,
that is, infinitesimally close to zero,
because $\hat{A}'$ and $\hat{v}'_1$ are
infinitesimally close to being orthogonal.
The same argument states that if
$f(\hat{A}')=1$, the probability of $f(\hat{v}'_2)=1$
must be infinitesimally close to zero.
Therefore, if $f(\hat{A}')=1$,
then $f(\hat{v}'_1)= f(\hat{v}'_2) =0$
for almost every system of an ensemble.
Using the same reasoning,
if $f(\hat{B}')=1$,
then $f(\hat{v}'_3)= f(\hat{v}'_4) =0$
for almost every system of an ensemble.
In addition,
if $f(\hat{v}'_1)=f(\hat{v}'_3)=0 \Rightarrow f(\hat{v}'_5)=1$ and
if $f(\hat{v}'_2)=f(\hat{v}'_4)=0 \Rightarrow f(\hat{v}'_6)=1$.
Since $f$ must simulate the predictions of QM,
if $f(\hat{v}'_5)=1$, then the probability of
$f(\hat{v}'_6)=1$ must be
$\left|{\langle {\hat{v}'_6|\hat{v}'_5} \rangle}\right|^2$,
that is, infinitesimally close to zero,
because $\hat{v}'_5$ and $\hat{v}'_6$ are
infinitesimally close to being orthogonal \cite{com6}.
\hfill\endproof


{\em Lemma 4.}
There is no KS coloring (\ref{sumrule})
of ${\cal D} \in S^2$ compatible
with all the statistical predictions of QM.

{\em Proof.}
I will use the same notation used in the proof of Lemma 2.
According to Lemma 3, in any NCHV theory
\begin{equation}
P(A' \wedge B') \approx 0.
\end{equation}
However, according to QM,
\begin{equation}
\left|{\langle {\hat{A}'|\hat{B}'} \rangle}\right|^2 \approx
\frac{1}{9}.
\end{equation}
\hfill\endproof


\section{Conclusion}


The reason why dense sets in \cite{Meyer99,Kent99,CK00} do not
lead to NCHV theories that simulate QM can be summarized as
follows: most of the many possible KS colorings of these sets must
be statistically irrelevant in order to reproduce some of the
statistical predictions of QM. Then, the remaining statistically
relevant KS colorings cannot reproduce some different statistical
predictions of QM. I therefore conclude that the existence of
KS-colorable sets that are dense in the corresponding Hilbert
spaces, like those in \cite{Meyer99,Kent99,CK00}, does not lead by
itself to a NCHV theory capable of eluding statistical KS-type
proofs, and must therefore not be interpreted as a nullification
of the physical impact of the KS theorem once the finite-precision
of measurements is taken into account.


\begin{acknowledgments}
The author thanks
D. M. Appleby,
R. Clifton,
C. A. Fuchs,
G. Garc\'{\i}a de Polavieja,
A. Kent,
A. J. L\'{o}pez Tarrida,
F. Mart\'{\i}n Maroto,
N. D. Mermin,
D. A. Meyer,
A. Peres,
E. Santos,
C. Serra,
C. Simon, and
M. \.{Z}ukowski
for discussions on this topic
and acknowledges support from
the organizers of the Sixth Benasque Center for Physics,
the University of Seville Grant No.~OGICYT-191-97,
the Junta de Andaluc\'{\i}a Grant No.~FQM-239, and
the Spanish Ministerio de Ciencia y Tecnolog\'{\i}a
Grant Nos.~BFM2000-0529 and BFM2001-3943.
\end{acknowledgments}



\begin{thebibliography}{99}
\bibitem{com2} If we were not
restricted to rational vectors,
then the maximum contradiction would
occur when $F_{\rm{QM}}=1.172$, as can be
numerically found.
Moreover, if we were not restricted to rational vectors,
we could choose $\hat{B}=\hat{C}$
while preserving the remaining orthogonalities
and then the maximum contradiction would occur when $F_{\rm{QM}}=\frac{1}{9}+1$
(see Refs.~\cite{Stairs83,Clifton93,Helle94,Vermaas94,CG95}).
\end{thebibliography}
\end{document}